\documentstyle[preprint,aps]{revtex}

\begin{document}

\draft

\preprint{September, 1999}

\title{Genuine Dyons in Born-Infeld Electrodynamics }

\author{Hongsu Kim}

\address{Department of Astronomy and Atmospheric Sciences \\
Kyungpook National University, Taegu, 702-701, KOREA \footnote{mailing address,
e-mail : hongsu@sirius.kyungpook.ac.kr} \\
and \\
Asia Pacific Center for Theoretical Physics\\
207-43 Cheongryangri-dong Dongdaemun-gu,
Seoul, 130-012, KOREA}

%\date{September, 1999}

\maketitle

\begin{abstract}
Study of magnetic monopoles in the original version of Born-Infeld (BI)
electrodynamics is performed. It then is realized that interesting new physics 
emerge and they include exotic behavior of radial electric monopole field 
such as its regularity as $r\rightarrow 0$ and its changing behavior with 
the absence or presence of the radial magnetic monopole field. This last 
point has been interpreted as the manifestation of the existence of point-like 
dyons in abelian BI theory. Two pieces of clear evidences in favor of this dyon
interpretation are provided. It is also demonstrated that despite these
unique features having no analogue in standard Maxwell theory, the
cherished Dirac quantisation condition remains unchanged. Lastly, comments
are given concerning that dyons found here in the original version of BI
electrodynamics should be distinguished from the ones with the same name or
BIons being studied in the recent literature on D-brane physics. 
\end{abstract}


\narrowtext
%\twocolumn

{\bf I. Introduction}
\\
Recently, the Born-Infeld (BI) theory has received much attention since the BI-type Lagrangians
naturally appear in string theories. Namely, it has been realized that they can describe the 
low energy dynamics of D-branes [2]. And this state of affair triggered the revival of interest
in the original BI electromagnetism [1] and further the exploration of BI gauge theories [2] 
in general.
Indeed, in spite of its long history this theory has remained almost unnoticed and hence 
nearly uncovered. This theory may be thought of as a highly nonlinear generalization of or
a non-trivial alternative to the standard Maxwell theory of electromagnetism. It is known that
Born and Infeld had been led, when they first constructed this theory, by the considerations
such as ; finiteness of the energy in electrodynamics, natural recovery of the usual Maxwell
theory as a linear approximation and the hope to find soliton-like solutions representing
point-like charged particles. In this respect, it seems that this theory, aside from its
connection to the recently fashionable brane physics [2], deserves serious and full exploration
in modern field theory perspective. And it is precisely this line of thought that initiated
the present study. Namely in this work, we would like to perform the study of magnetic monopole
in the original version of pure abelian BI gauge theory [1]. To be a little more concrete, we shall
introduce the magnetic charge current density in BI equations just as Dirac did in Maxwell
equations and see if this introduction can provide the BI equations with dual symmetry. 
Although this turns out not to be the case, we find that the abelian BI gauge theory exhibits
unique features which have no analogues in the standard Maxwell theory. That is, due to the
lack of dual symmetry, the static electric monopole field and the static magnetic monopole
field have different $r$-dependences. To be more specific, the static electric monopole field
shows exotic behaviors such as the regularity as $r\rightarrow 0$ signaling the finiteness
of the energy stored in the field of electric point charge. When both electric and magnetic
monopoles are present (and are located, say, at the same point), surprises continue and
particularly the behavior of electric monopole field in the presence of the magnetic monopole
becomes different from that in the absence of the magnetic monopole. Obviously, this can be
attributed to the highly non-linear nature of the BI theory action which then results in the
direct and unique coupling between the electric and magnetic fields in a highly non-trivial
manner even in the static case and hence is a really unique feature having no analogue in the 
usual Maxwell theory. We then interpret this exotic behavior of monopole fields in the presence 
of both electric and magnetic charges arising from the unique coupling between electric and 
magnetic fields as the manifestation of the existence of ``dyons'' even in abelian BI gauge theory.
Indeed two pieces of clear evidences in favor of the dyon interpretation will be provided and
one of which employs the argument based on the translations of monopole fields and the other 
invokes the energetics argument. Finally, we shall point out that despite all theses unique
features of monopoles and dyons in BI electrodynamics, something seems never change and that is
the cherished Dirac quantisation condition and the meaning that underlies it.   
One might wonder how the dyon solutions found in this work should be understood in relation to 
the soliton solutions dubbed ``BIons'' or just dyons studied in the recent literature on brane
physics [3-9]. Thus later on in the concluding remarks section, we shall comment on this point in detail.
\\
{\bf II. Monopole fields in Maxwell gauge theory}
\\
As stated, our main objective in this work is the parallel study of physics of magnetic monopoles
between in abelian BI electrodynamics and in the standard Maxwell electrodynamics. Thus to this
end, we begin with the brief review of Dirac's proposal for the introduction of magnetic
monopoles in Maxwell theory. Consider the action for the Maxwell gauge theory in flat spacetime
(and in MKS unit)
\begin{eqnarray}
S = \int d^4x \left[ -{1\over 4}F_{\mu\nu}F^{\mu\nu} + j^{\mu}A_{\mu} \right]
\end{eqnarray}
where $F_{\mu\nu}=\partial_{\mu}A_{\nu}-\partial_{\nu}A_{\mu}$ is the field strength 
and $j^{\mu}= (\rho_{e}, ~j^{i}_{e})$ is the electric source current for the abelian 
field $A_{\mu}$. Extremizing this action with respect to the gauge field $A_{\mu}$, then,
yields the field equation for $A_{\mu}$ as 
\begin{eqnarray}
\partial_{\mu}F^{\mu\nu} = - j^{\nu}.
\end{eqnarray} 
In addition to this, there is a supplementary equation coming from an identity satisfied by
the abelian gauge field strength tensor, 
$\partial_{\lambda}F_{\mu\nu} + \partial_{\mu}F_{\nu\lambda} + \partial_{\nu}F_{\lambda\mu}=0.$
This is the Bianchi identity which is just a geometrical equation and in terms of the Hodge 
dual field strength, $\tilde{F}^{\mu\nu} = {1\over 2}\epsilon^{\mu\nu\alpha\beta}F_{\alpha\beta}$,
it can be written as
\begin{eqnarray}
\partial_{\mu}\tilde{F}^{\mu\nu} = 0
\end{eqnarray}
And it seems noteworthy that the field equation for $A_{\mu}$ in eq.(2) is the dynamical field
equation which gets determined by the concrete nature of the gauge theory action such as the
one in eq.(1). The Bianchi identity in eq.(3), on the other hand, is simply a geometrical
identity and is completely independent of the choice of the context of the gauge theory.
This set of four equations, known as the Maxwell equations for classical electrodynamics,
however, may be viewed as being somewhat incomplete in that the right hand side (r.h.s) of the Bianchi
identity is vacant. Thus Dirac proposes to make it look complete  
by adding the ``magnetic current density'' term $k^{\mu} = (\rho_{m}, ~j^{i}_{m})$ on the r.h.s.
Further, if one decomposes these covariant equations using $\partial^{\mu} = (-\partial/\partial t,
\nabla_{i})$, $\partial_{\mu}=\eta_{\mu\nu}\partial^{\nu} = (\partial/\partial t, \nabla_{i})$
(namely, we use the sign convention, $\eta_{\mu\nu} = diag(-1, 1, 1, 1)$), $A^{\mu}=(\phi, A^{i})$
and the field identification, $E_{i}=F_{i0}$, $B_{i}={1\over 2}\epsilon_{ijk}F^{jk}$ or
$F_{ij}=\epsilon_{ijk}B^{k}$, the dynamical field equation decomposes into
\begin{eqnarray}
\nabla \cdot \vec{E} = \rho_{e},  ~~~\nabla \times \vec{B} - {\partial \vec{E}\over \partial t}
= \vec{j}_{e}
\end{eqnarray}
while the geometrical Bianchi identity decomposes as 
\begin{eqnarray}
\nabla \cdot \vec{B} = \rho_{m},  ~~~\nabla \times \vec{E} + {\partial \vec{B}\over \partial t} 
= - \vec{j}_{m}. 
\end{eqnarray}
These Maxwell equations are then invariant under the ``duality transformation'' (here in this work,
we restrict ourselves to the discrete duality transformation, {\it not} the continuous duality
rotations [4]) 
\begin{eqnarray}
F^{\mu\nu} \rightarrow \tilde{F}^{\mu\nu} (\vec{E} \rightarrow \vec{B}),   
~~~j^{\mu}\rightarrow k^{\mu} \\
{\rm and} ~~~\tilde{F}^{\mu\nu} \rightarrow - F^{\mu\nu} (\vec{B}  \rightarrow - \vec{E}),
~~~k^{\mu} \rightarrow - j^{\mu}. \nonumber
\end{eqnarray}
Note, however, that this dual invariance is just a symmetry in the classical field equation and
the Bianchi identity but {\it not} in the Maxwell theory action
$L = - {1\over 4}F_{\mu\nu}F^{\mu\nu}$ as $\tilde{F}_{\mu\nu}\tilde{F}^{\mu\nu} = - 
F_{\mu\nu}F^{\mu\nu}$. Lastly, before we end our review of magnetic monopoles in Maxwell gauge
theory, we recall, for later use, that the expressions for the static electric and magnetic fields 
generated by electric and magnetic monopoles sitting at the origin and hence the solutions to
$\nabla \cdot \vec{E} = e\delta^3(\vec{r})$ and $\nabla \cdot \vec{B} = g\delta^3(\vec{r})$
are given respectively by 
$\vec{E} = (e/4\pi r^2)\hat{r}$ and  $\vec{B} = (g/4\pi r^2)\hat{r}$ for $r\neq 0$.
They have the same structure, i.e., the identical $r$-dependences as can be expected from the 
dual symmetry of the Maxwell equations. This point, which is so familiar and hence looks trivial,
will be contrasted to what happens in the study of static monopole fields in abelian BI
gauge theory to which we now turn. 
\\
{\bf III. Monopole and Dyon fields in abelian BI gauge theory}
\\
As usual, we begin with the action for this abelian BI theory
which is given, in 4-dimensions, by [2]
\begin{eqnarray}
S &=& \int d^4x \left\{ \beta^2\left[1 - \sqrt{-det(\eta_{\mu\nu} + 
{1\over \beta}F_{\mu\nu})}\right]
+ j^{\mu}A_{\mu}\right\} \\
&=& \int d^4x \left\{ \beta^2\left[1 - \sqrt{1 + {1\over 2\beta^2}F_{\mu\nu}F^{\mu\nu} -
{1\over 16\beta^4}(F_{\mu\nu}\tilde{F}^{\mu\nu})^2}\right] + j^{\mu}A_{\mu}\right\} \nonumber
\end{eqnarray}
where ``$\beta$'' is a generic parameter of the theory having the dimension $dim[\beta] =
dim[F_{\mu\nu}] = +2$. It probes the degree of deviation of BI gauge theory from the standard 
Maxwell theory and obviously $\beta \rightarrow \infty$ limit corresponds to the standard
Maxwell theory. Again, extremzing this action with respect to $A_{\mu}$ yields the dynamical
BI field equation 
\begin{eqnarray}
\partial_{\mu}\left[{{F^{\mu\nu} - {1\over 4\beta^2}(F_{\alpha\beta}\tilde{F}^{\alpha\beta})
\tilde{F}^{\mu\nu}} \over {\sqrt{1 + {1\over 2\beta^2}F_{\alpha\beta}F^{\alpha\beta} - 
{1\over 16\beta^4}(F_{\alpha\beta}\tilde{F}^{\alpha\beta})^2}}}\right] = - j^{\nu}.
\end{eqnarray}
The geometrical Bianchi identity, which is a supplementary equation to this field equation is,
as mentioned earlier, independent of the nature of the gauge theory action. Thus it is
\begin{eqnarray}
\partial_{\mu}\tilde{F}^{\mu\nu} = 0.
\end{eqnarray}
As before, we now split up these covariant equations and write them in terms of $\vec{E}$ and
$\vec{B}$ fields to get
\begin{eqnarray}
&\nabla & \cdot \left[{1\over R}(\vec{E} + {1\over \beta^2}(\vec{E}\cdot \vec{B})\vec{B})\right]
= \rho_{e}, \\
&\nabla & \times \left[{1\over R}(\vec{B} - {1\over \beta^2}(\vec{E}\cdot \vec{B})\vec{E})\right] 
- {\partial \over \partial t}
\left[{1\over R}(\vec{E} + {1\over \beta^2}(\vec{E}\cdot \vec{B})\vec{B})\right] = \vec{j}_{e}
\nonumber
\end{eqnarray}
where $R \equiv \sqrt{1 + {1\over 2\beta^2}F_{\alpha\beta}F^{\alpha\beta} -
{1\over 16\beta^4}(F_{\alpha\beta}\tilde{F}^{\alpha\beta})^2} = \sqrt{1 - {1\over \beta^2}
(\vec{E}^2 - \vec{B}^2) - {1\over \beta^4}(\vec{E}\cdot \vec{B})^2}$ for the dynamical BI field
equation and 
\begin{eqnarray}
\nabla \cdot \vec{B} = 0,  ~~~\nabla \times \vec{E} + {\partial \vec{B}\over \partial t} = 0
\end{eqnarray}
for the geometrical Bianchi identity and where we used $F_{\mu\nu}F^{\mu\nu} = -2(\vec{E}^2-\vec{B}^2)$
and $F_{\mu\nu}\tilde{F}^{\mu\nu} = 4\vec{E}\cdot \vec{B}$. 
We now consider the case when both the electric current density $j^{\mu} = (\rho_{e}, \vec{j}_{e})$
and the magnetic current density $k^{\mu} = (\rho_{m}, \vec{j}_{m})$ are present.
Then as before, the Bianchi identity gets modified to
\begin{eqnarray}
\nabla \cdot \vec{B} = \rho_{m},  ~~~\nabla \times \vec{E} + {\partial \vec{B}\over \partial t}
= - \vec{j}_{m}
\end{eqnarray}
or, in covariant form, to $\partial_{\mu}\tilde{F}^{\mu\nu} = - k^{\nu}$.
One can then readily realize that, unlike the Maxwell equations, these four BI equations evidently
do not possess the duality invariance mentioned earlier. Namely, the two dynamical BI field
equations and the remaining two geometrical Bianchi identity are not dual to each other any more
and it can be attributed to the fact that when passing from the standard Maxwell to this highly
nonlinear BI theory, only the dynamical field equations experience non-trivial change 
(``nonlinearization'') and the geometrical Bianchi identity remains unchanged as it is independent
of the nature of gauge theory itself. This inherent lack of dual invariance in BI equations then
may imply that we need not introduce the point-like magnetic monopole and current in the first
place. But for the sake of parallel study of the interesting monopole physics
in Maxwell theory, here we shall assume the existence of point-like magnetic monopole and explore
the physics of it such as the structure of static monopole fields and the possible existence of
the dyon solution. First, we consider the static magnetic monopole field and electric monopole 
field in this abelian BI gauge theory. The static magnetic monopole field can be obtained by
solving one of the Bianchi identity equations, $\nabla \cdot \vec{B} = g\delta^3(\vec{r})$. For
$\vec{r}\neq 0$, and in spherical-polar coordinates, this equation is given by
$[\partial_{r}(r^2 \sin\theta B_{r}) + \partial_{\theta}(r \sin\theta B_{\theta}) +
\partial_{\phi}(r B_{\phi})]/r^2\sin\theta = 0$ and is solved by 
$B_{r}(r) = g/4\pi r^2$, $B_{\theta} = B_{\phi} = 0$. Note that this solution form holds 
irrespective of the existence of the electric monopole. Next, the static electric monopole 
field can be obtained from one of the dynamical field equations
$\nabla \cdot [\left\{\vec{E}+(\vec{E}\cdot \vec{B})\vec{B}/\beta^2\right\}/R] = e\delta^3(\vec{r})$
with $R \equiv \left\{1 - (\vec{E}^2-\vec{B}^2)/\beta^2 - (\vec{E}\cdot \vec{B})^2/\beta^4\right\}^{1/2}$.
Again for $\vec{r}\neq 0$, and in spherical-polar coordinates, this equation becomes
$[\partial_{r}(r^2 \sin\theta \hat{E}_{r}) + \partial_{\theta}(r \sin\theta \hat{E}_{\theta}) +
\partial_{\phi}(r \hat{E}_{\phi})]/r^2\sin\theta = 0$ with
$\hat{E}_{i} \equiv [E_{i}+(\vec{E}\cdot \vec{B})B_{i}/\beta^2]/R$. Firstly, in the absence of the
magnetic monopole, $\hat{E}_{i} = E_{i}/\sqrt{1-\vec{E}^2/\beta^2}$ and then the above equation is
solved by $E_{r}(r) = e/4\pi \sqrt{r^4 + (e/4\pi\beta)^2}$, $E_{\theta} = E_{\phi} = 0$. Next, in the
presence of the magnetic monopole, one has to put the magnetic monopole field 
$\vec{B} = (g/4\pi r^2)\hat{r}$ in $\hat{E}_{i}$ and $R$ and then solve the equation. Then the equation
admits the solution $E_{r}(r) = e/4\pi \sqrt{r^4 + (e^2+g^2)/(4\pi\beta)^2}$,$E_{\theta} = E_{\phi} = 0$.
The static magnetic monopole field in the abelian BI theory, therefore, turns out to be identical to 
that in the standard Maxwell theory. Concerning the static electric monopole field in this abelian BI
theory, however, there are two peculiar features worthy of note. For one thing, unlike in the Maxwell
theory, the electric monopole field and the magnetic monopole field exhibit different $r$-dependences
which can be attributed to the fact that in this BI theory, the dynamical field equation and the Bianchi
identity are not dual to each other. Besides, since the static electric monopole field is not singular
as $r \rightarrow 0$, the energy stored in the field of electric point charge could be finite and this
point seems to be consistent with the consideration of finiteness of energy, which is one of the
motivations to propose this BI electrodynamics when it was first devised. For the other, it is very
interesting to note that the static electric monopole field gets modified when the magnetic monopole
(field) is present although the $r$-dependence itself remains essentially the same as we observed above.
This is indeed a very peculiar feature which is unique and has no analogue in the standard Maxwell theory.
And when the magnetic monopole (field) is present, the strength of the static electric monopole field
appears to experience some attenuation which is particularly noticeable in the ``small-$r$'' region,
when compared to the case without the magnetic monopole field. It is also tempting to interpret this
unique coupling between the electric and magnetic field even in the static monopole case (which evidently
originates from the highly nonlinear nature of BI theory action in 4-dimensions) as the manifestation of
the existence of ``dyon'' even in ``abelian'' BI gauge theory. Thus we elaborate on this particularly
interesting point. First notice that the evaluation of the $det(\eta_{\mu\nu}+F_{\mu\nu}/\beta)$ in
the general form of the BI action particularly in 4-dimensions produces the term
$(F_{\mu\nu}\tilde{F}^{\mu\nu})^2/16\beta^4 = (\vec{E}\cdot \vec{B})^2/\beta^4$, i.e.,
\begin{eqnarray}
\sqrt{ - det(\eta_{\mu\nu} + {1\over \beta}F_{\mu\nu})} = \sqrt{ 1 + {1\over 2\beta^2}F_{\mu\nu}F^{\mu\nu}
 - {1\over 16\beta^4}(F_{\mu\nu}\tilde{F}^{\mu\nu})^2}. \nonumber
\end{eqnarray}
And it is precisely this term which induces a unique and direct coupling between the electric and magnetic
field even in the static monopole case and hence generates the dyon solution. 
Then next, we seem to be left with the question :
what are the evidences that would support the dyon interpretation of monopole solutions ?
\begin{eqnarray}
\vec{B} = {g\over 4\pi r^2}\hat{r},
~~~\vec{E} = {e\over {4\pi \sqrt{r^4 + {e^2+g^2\over (4\pi \beta)^2}}}}\hat{r}.
\end{eqnarray}
Even in a loose sense, two pieces of evidences can be suggested. We begin with the first one.
Consider that in the standard Maxwell theory, the monopole solutions
$\vec{B}=\left\{g/4\pi |\vec{r}-\vec{r}_{B}|^3\right\}(\vec{r}-\vec{r}_{B})$ and
$\vec{E}=\left\{e/4\pi |\vec{r}-\vec{r}_{E}|^3\right\}(\vec{r}-\vec{r}_{E})$, corresponding
to the configuration in which the magnetic charge $g$ is fixed at $\vec{r}=\vec{r}_{B}$ and
the electric charge is fixed at $\vec{r}=\vec{r}_{E}$ separately, are simultaneous solutions 
to the Maxwell equations. And the particular solutions $\vec{B} = \left\{g/4\pi r^2\right\}\hat{r}$
and $\vec{E} = \left\{e/4\pi r^2\right\}\hat{r}$ just represent the case when the two monopoles
$g$ and $e$ happen to be sitting on the same location, the origin. In the BI theory of electromagnetism,
however, the particular solutions given in eq.(13) above do not simply represent the case when the
two monopole, $g$ and $e$ are sitting separately at the origin. Instead, these particular static
monopole solutions represent a single, point-like entity carrying both electric and magnetic charges,
i.e., the ``point-like dyon''. To see if this is indeed the case, notice that
$\vec{B}=\left\{g/4\pi |\vec{r}-\vec{r}_{B}|^2\right\}(\hat{r}-\hat{r}_{B})$ and
$\vec{E}=\left\{e/4\pi \sqrt{|\vec{r}-\vec{r}_{E}|^4 + [e^2+g^2/(4\pi \beta)^2]}\right\}
(\hat{r}-\hat{r}_{E})$ (where $(\hat{r}-\hat{r}_{B})\equiv (\vec{r}-\vec{r}_{B})/|\vec{r}-\vec{r}_{B}|$)
fail to be simultaneous solutions to the BI equations
$\nabla \cdot \vec{B} = g\delta^3(\vec{r}-\vec{r}_{B})$ and
$\nabla \cdot [\left\{\vec{E}+(\vec{E}\cdot \vec{B})\vec{B}/\beta^2\right\}/R] =
e\delta^3(\vec{r}-\vec{r}_{E})$
(where again $R \equiv \left\{1 - (\vec{E}^2-\vec{B}^2)/\beta^2 - 
(\vec{E}\cdot \vec{B})^2/\beta^4\right\}^{1/2}$)
for $\vec{r}_{B} \neq \vec{r}_{E}$. They, however, can be simultaneous solutions only for
$\vec{r}_{B} = \vec{r}_{E}$, namely only when $g$ and $e$ stick to each other. And it is straightforward to
check that the static monopole solutions to BI equations for $\vec{r}_{B} \neq \vec{r}_{E}$, when actually
worked out, turn out to take different structures from those given above by simply replacing 
$\vec{r}\rightarrow (\vec{r}-\vec{r}_{B})$ and $\vec{r}\rightarrow (\vec{r}-\vec{r}_{E})$. 
This is certainly in sharp contrast to what happens in standard Maxwell electromagnetism where 
$\vec{B}=\left\{g/4\pi |\vec{r}-\vec{r}_{B}|^2\right\}(\hat{r}-\hat{r}_{B})$ and
$\vec{E}=\left\{e/4\pi |\vec{r}-\vec{r}_{E}|^2\right\}(\hat{r}-\hat{r}_{E})$, which are obtained simply by
replacing $\vec{r}\rightarrow (\vec{r}-\vec{r}_{B})$ and $\vec{r}\rightarrow (\vec{r}-\vec{r}_{E})$ 
are legitimate and unique
solutions to the Maxwell equations even for $\vec{r}_{B} \neq \vec{r}_{E}$. Undoubtedly, this observation
implies that in BI theory, when both electric and magnetic charges are present, they can stick together
to form a point-like dyon and the static electric and magnetic fields it produces are given by the expressions
given above with $\vec{r}_{B} = \vec{r}_{E}$. Thus this can be thought of as one clear evidence in favor of
the dyon interpretation and the other can be derived in terms of energetics (argument in terms of energy)
as follows. Consider the energy-momentum tensor of this BI theory
\begin{eqnarray}
T_{\mu\nu} = \beta^2 (1 - R)\eta_{\mu\nu} + {1\over R}\left[F_{\mu\alpha}F_{\nu}^{\alpha} -
{1\over 4\beta^2}(F_{\alpha\beta}F^{\alpha\beta})F_{\mu\alpha}\tilde{F}_{\nu}^{\alpha}\right]
\end{eqnarray}
with $R$ as given earlier. The energy density stored in the electromagnetic field can then be read off
as
\begin{eqnarray}
T_{00} = \beta^2 \left[{1 + {1\over \beta^2}\vec{B}^2 \over {\sqrt{1 - {1\over \beta^2}
(\vec{E}^2 - \vec{B}^2) - {1\over \beta^4}(\vec{E}\cdot \vec{B})^2}}} - 1 \right]
\end{eqnarray}
which does reduce to its Maxwell theory's counterpart $(\vec{E}^2 + \vec{B}^2)/2$ in the limit
$\beta \rightarrow \infty$ as it should. We now compute the energy density solely due to the 
magnetic field generated by the magnetic charge $g$. Using $\vec{B} = (g/4\pi r^2)\hat{r}$,
\begin{eqnarray}
T^{B}_{00} = \beta^2 \left[\sqrt{1 + {1\over \beta^2}\vec{B}^2} - 1\right] =
\beta^2 \left[\sqrt{1 + {g^2\over (4\pi\beta)^2}{1\over r^4}} - 1\right].
\end{eqnarray}
Next, we calculate the energy density stored in the electric field generated by the electric charge $e$.
Then using $\vec{E} = \left\{e/4\pi \sqrt{r^4 + (e/4\pi\beta)^2}\right\}\hat{r}$, one gets
\begin{eqnarray}
T^{E}_{00} = \beta^2 \left[{1\over \sqrt{1 - {1\over \beta^2}\vec{E}^2}} - 1\right] =
\beta^2 \left[\sqrt{1 + {e^2\over (4\pi\beta)^2}{1\over r^4}} - 1\right].
\end{eqnarray}
There now seem to be two points worthy of note. One is the fact that $T^{B}_{00}$ and $T^{E}_{00}$ are
basically the same except that $g$ and $e$ are interchanged although the magnetic monopole field $\vec{B}$
and the electric monopole field $\vec{E}$ possess different $r$-dependences. The other is, as Born and
Infeld hoped when they constructed this new theory, the energy stored in a static monopole field is 
indeed finite. For instance, the electric monopole energy can be evaluated in a concrete manner as [4] 
\begin{eqnarray}
E &=& \int d^3x T^{E}_{00} = \int^{\infty}_{0}dr \beta^2 \left[\sqrt{(4\pi r^2)^2 + {e^2\over \beta^2}}
- 4\pi r^2\right] \nonumber \\
&=& \sqrt{{\beta e^3\over 4\pi}}\int^{\infty}_{0}dy \left[\sqrt{y^4 + 1} - y^2\right] \\
&=& \sqrt{{\beta e^3\over 4\pi}}{\pi^{3/2}\over 3\Gamma({3\over 4})^2} =
1.23604978 \sqrt{{\beta e^3\over 4\pi}} \nonumber 
\end{eqnarray}
where $y^2 = (4\pi \beta/e^2)r^2$ and in the $y$-integral, integration by part and the elliptic
integral have been used. Coming back to the argument based on the energetics, lastly we compute
the energy density due to the electric and magnetic fields generated by both the electric charge
and the magnetic charge. Substituting the expressions given in eq.(13) into eq.(15), one gets
\begin{eqnarray}
T^{E+B}_{00} = \beta^2 \left[{{1 + {g^2\over (4\pi \beta)^2}{1\over r^4}}\over
\sqrt{1 - {1\over (4\pi \beta)^2}{1\over \left\{r^4+{e^2+g^2\over (4\pi\beta)^2}\right\}}
\left[e^2 - g^2\left\{1 + {g^2\over (4\pi \beta)^2}{1\over r^4}\right\}\right]}} - 1\right].
\end{eqnarray}
Now note the following two points : concerning the dyon interpretation, we would like to draw some
hints by comparing $T^{E+B}_{00}$ with $T^{E}_{00}+T^{B}_{00}$. Firstly, $T^{E+B}_{00}$ turns out not
to be symmetric under $e\leftrightarrow g$ whereas $T^{E}_{00}+T^{B}_{00}$ was as we noticed earlier.
Secondly, the sum of energy density of magnetic monopole field alone $T^{B}_{00}$ and energy density
of electric monopole field alone $T^{E}_{00}$ is {\it not} equal to the energy density of electric
and magnetic fields when both the electric and magnetic monopoles are present, i.e.,
$T^{E}_{00}+T^{B}_{00}\neq T^{E+B}_{00}$. Again these observations are in apparent contrast to what 
happens in standard Maxwell theory where
$T^{E}_{00}+T^{B}_{00} = {1\over 2}(\vec{E}^2 + \vec{B}^2) = {e^2+g^2\over 2(4\pi)^2}({1\over r^4})
 = T^{E+B}_{00}$ and hence the possibility of $e-g$ bound state or dyon is completely excluded even
if they are forced to be brought together. Here, when considering the total energy of $e-g$ system,
one might wonder why the $e-g$ interaction potential energy is not taken into account. Recall,
however, that we are considering only the ``static'' case when both $e$ and $g$ are held fixed at
each position and hence exert no force to each other. To see this, note that the Lorentz force law
is generalized in the presence of both electric and magnetic charges to (the validity of this
Lorentz force law even in the context of BI electrodynamics will be discussed carefully later on)
\begin{eqnarray}
m{d^2x^{\mu}\over d\tau^2} = \left(eF^{\mu\nu} + g\tilde{F}^{\mu\nu}\right){dx_{\nu}\over d\tau}
\nonumber
\end{eqnarray}
or in components, to 
\begin{eqnarray}
\vec{F} = e(\vec{E} + \vec{v}\times \vec{B}) + g(\vec{B} - \vec{v}\times \vec{E}) 
\end{eqnarray}
from which one can realize that, unlike the homogeneous systems of electric charges alone or magnetic
charges alone, the interaction force (and hence the potential energy) between $e$ and $g$ arises only
when one of the two is in motion relative to the other. Therefore in the static case, there is no interaction 
force and potential energy between static $e$ and $g$ and thus the total energy density of $e-g$ 
system is given simply by $T^{E+B}_{00}$.
Therefore this consideration of energetics of $e-g$ system also appears to provide another concrete
evidence in favor of the existence of a $e-g$ bound state, i.e., a dyon in BI theory of electromagnetism
although the definite statement can be made if one could somehow show $T^{E+B}_{00} < 
T^{E}_{00}+T^{B}_{00}$
which, at least to us, does not look so easy to demonstrate in a straightforward manner. 
Lastly, one can realize that despite all these unique and interesting features, the static monopole fields
in non-trivial BI theory (i.e., for ``finite''-$\beta$) are effectively indistinguishable from those in the
standard Maxwell theory in the far ($r\rightarrow \infty$) zone and the possibly significant deviations
occur only in the near ($r\rightarrow 0$) zone. \\
Before we close the study of monopoles and dyons in abelian BI gauge theory, we would like to make one more point
which seems worthy of note. In the standard Maxwell theory, the motion of an electrically charged particle 
in a radial magnetic monopole field is of some interest. Thus we now consider the motion of an
electrically
charged ``test'' particle in the external ``background'' magnetic field generated by a static magnetic monopole
acting as just a source. (Thus this situation should be distinguished from the system of static electric charge
$e$ and magnetic charge $g$ we considered above when both $e$ and $g$ are fixed at each position and thus treated
as sources for static monopole fields.) To be a little more specific, it is well-known that in this system the 
conserved quantity is not just the orbital angular momentum of the charged test particle, $\vec{L}$, but the ``total''
angular momentum given by 
\begin{eqnarray}
\vec{J} = \vec{L} - {eg\over 4\pi}\hat{r}
\end{eqnarray}
where $\vec{J}_{em} = \int d^3x [\vec{x}\times(\vec{E}\times \vec{B})] = -(eg/4\pi)\hat{r}$ is the angular
momentum of the charged point particle's electric field obeying $\nabla\cdot \vec{E} = e \delta^3(\vec{x}-
\vec{r})$ (with $\vec{r}$ being the trajectory of the electric charge) and the static monopole's magnetic
field $\vec{B} = (g/4\pi r^2)\hat{x}$. In quantized version of the theory, then, one expects components of
$\vec{J}$ to satisfy the usual angular momentum commutation relations implying that the eigenvalues of
$J_{i}$ are half integers. Since the orbital angular momentum $\vec{L}$ is expected to have integral
eigenvalues, one then gets $eg/4\pi = n/2$ with $n$ being integers. Thus eq.(21), in turn, implies the Dirac
quantization condition 
\begin{eqnarray}
eg = 2\pi n.
\end{eqnarray}
Then one might wonder what would happen to the same test electric charge-source monopole system particularly
concerning the Dirac quantization rule in the context of abelian BI theory. The answer is, interestingly,
that no essential changes occur. To see this, we first attempt to derive the expression for the conserved
total angular momentum of this system. To do so, however, one needs to know the BI theory version of Lorentz
force law. As we mentioned earlier, indeed the Lorentz force law is determined in a gauge theory-independent
manner. This can be readily checked as follows. We begin with the 4-vector current of a charged particle 
localized on its spacetime trajectory $x^{\mu}(\tau)$ with $\tau$ being particle's proper time
\begin{eqnarray}
j^{\mu}(t,\vec{y}) = e {dx^{\mu}\over dt}\delta^3 [\vec{y} - \vec{x}(\tau)]|_{t=x^{0}(\tau)} =
e\int d\tau {dx^{\mu}\over d\tau}\delta^4 [y - x(\tau)]
\nonumber
\end{eqnarray}
which fulfills the continuity equation $\partial_{\mu}j^{\mu}=0$.
Now to see what the Lorentz force law would look like in the context of BI electrodynamics, we consider
the combined system of a charged test particle and a given background gauge field in abelian BI theory
described by the action
\begin{eqnarray}
S &=& \int d^4x\left[{\it L}_{BI} + j^{\mu}A_{\mu}\right] - m\int ds \nonumber \\
&=& \int d^4x {\it L}_{BI} + e\int dx^{\mu}A_{\mu}[x(\tau)] - m\int ds \\
&=& \int d^4x {\it L}_{BI} + \int dt \left[-m\sqrt{1 - v^2_{i}} - eA^{0} + eA^{i}v_{i}\right] \nonumber 
\end{eqnarray}
where ${\it L}_{BI}$ is the abelian BI theory action given earlier and we used 
$ds = d\tau = dt\sqrt{1 - v^2_{i}}$. Apparently, any charged particle acts as an additional source thus
modifying the surrounding field. If, however, we neglect this ``back reaction'' effect as a first
approximation and assume $A^{\mu}$ as just an external background field, we may leave out the gauge
field action as the external field serves as just a ``hard'' background. Thus in this usual
approximation in which the back reaction of the charged test particle to the surrounding field is
neglected, the motion of the charged particle becomes independent of the detailed dynamical nature 
of the gauge theory itself. Therefore the motion of a charged particle under the influence of a
given external gauge field would be governed by the action
\begin{eqnarray}
S = \int dt \left[-m\sqrt{1 - v^2_{i}} + e(A^{i}v_{i} - A^{0})\right] \nonumber
\end{eqnarray}
and by extremizing it with respect to $x^{i}$, one gets the following Euler-Lagrange's equation of
motion $d\vec{P}/dt = e(\vec{E} + \vec{v}\times \vec{B})$ where $P^{i} = mv^{i}/\sqrt{1 - v^2_{i}}$.
Since this is the usual Lorentz force law, we can realize that indeed it holds irrespective of the
context of the dynamical gauge theory as stated above. Thus even in this abelian BI gauge theory,
the rate of change of the orbital angular momentum of the system consisting of the electrically
charged test particle and the source magnetic monopole is
\begin{eqnarray}
{d\vec{L}\over dt} = \vec{r}\times {d\vec{P}\over dt} = {eg\over 4\pi r^3}\vec{r}\times
(\dot{\vec{r}}\times \vec{r}) = {d\over dt}\left({eg\over 4\pi}\hat{r}\right) \nonumber
\end{eqnarray}
again suggesting that the conserved total angular momentum is given by $\vec{J}=\vec{L}-
(eg/4\pi)\hat{r}$ just as in the standard Maxwell theory. And this result implies that the
usual Dirac quantization condition still holds true in abelian BI theory as well. In addition,
we also realized in this work that in this BI theory, point-like dyons as well as magnetic
monopoles can exist. And it is not hard to see that even in the combined system of an electrically
charged test particle and a static source dyon, the conserved total angular momentum is the same as 
in the test particle-monopole system and hence the Dirac quantization condition also remains the
same. Next, it seems worthy of note that the interpretation of this total angular momentum as the
sum of orbital angular momentum $\vec{L}$ of the test electric charge $e$ and the field angular
momentum $\vec{J}_{em} = - (eg/4\pi)\hat{r}$ due to the electric charge $e$ and the magnetic 
charge $g$ also stays the same as in the Maxwell theory case. Namely, the angular momentum is
passed back and forth between the electric charge and the field as it is expected to be. This
statement sounds natural and hence can be taken for granted. But to demonstrate that this is indeed
the case even in the context of BI theory is not so trivial and hence seems worth doing. Thus in
the following, we briefly sketch the demonstration. And to do so, we need some preparation.
In the dynamical BI field equations given earlier, we define, for the sake of convenience of the
formulation, the ``electric displacement'' $\vec{D}$ and the ``magnetic field'' $\vec{H}$ in terms of
the fundamental fields $\vec{E}$ and $\vec{B}$ as
\begin{eqnarray}
\vec{D} = {1\over R}\left\{\vec{E} + {1\over \beta^2}(\vec{E}\cdot \vec{B})\vec{B}\right\},
~~~\vec{H} = {1\over R}\left\{\vec{B} - {1\over \beta^2}(\vec{E}\cdot \vec{B})\vec{E}\right\}  
\nonumber
\end{eqnarray}
where $R = \sqrt{1 - {1\over \beta^2}(\vec{E}^2 - \vec{B}^2) - {1\over \beta^4}(\vec{E}\cdot \vec{B})^2}$
is as defined earlier. Then the BI equations take the form 
\begin{eqnarray}
\nabla \cdot \vec{D} &=& \rho_{e},  ~~~\nabla \times \vec{H} - {\partial \vec{D}\over \partial t}
= \vec{j}_{e} \\
\nabla \cdot \vec{B} &=& \rho_{m},  ~~~\nabla \times \vec{E} + {\partial \vec{B}\over \partial t}
= - \vec{j}_{m}. \nonumber
\end{eqnarray}
Now $\vec{E}\cdot ({\rm Ampere's ~law ~eq.}) - \vec{H}\cdot ({\rm Faraday's ~induction ~law ~eq.})$ yields
\begin{eqnarray}
\vec{H}\cdot (\nabla \times \vec{E}) - \vec{E}\cdot (\nabla \times \vec{H}) =
- \vec{H}\cdot {\partial \vec{B} \over \partial t} - \vec{E}\cdot {\partial \vec{D} \over \partial t}
- \vec{j}_{e}\cdot \vec{E} - \vec{j}_{m}\cdot \vec{H}. \nonumber
\end{eqnarray}
Further using
\begin{eqnarray}
\vec{H}\cdot (\nabla \times \vec{E}) - \vec{E}\cdot (\nabla \times \vec{H}) = \nabla \cdot 
(\vec{E}\times \vec{H}), 
~~~- \vec{H}\cdot {\partial \vec{B} \over \partial t} - \vec{E}\cdot {\partial \vec{D} \over \partial t}
= - {\partial \over \partial t}T_{00} \nonumber
\end{eqnarray}
where $T_{00}$ is the energy density stored in the electromagnetic field in BI theory given in eq.(15),
one arrives at the familiar local energy conservation equation
\begin{eqnarray}
\nabla \cdot \vec{S} + {\partial u \over \partial t} = - \vec{j}_{e}\cdot \vec{E} - \vec{j}_{m}\cdot \vec{H}
\end{eqnarray}
where $u = T_{00}$ is the energy density, $\vec{S} = \vec{E} \times \vec{H}$ is the ``Poynting vector''
representing the local energy flow per unit time per unit area and 
$- \vec{j}_{e}\cdot \vec{E} - \vec{j}_{m}\cdot \vec{H}$ on the right hand side is the power dissipation
per unit volume. In particular for $\vec{j}_{e}\cdot \vec{E} = 0 = \vec{j}_{m}\cdot \vec{H}$, one gets
\begin{eqnarray}
\nabla \cdot \vec{S} + {\partial u \over \partial t} = 0 \nonumber
\end{eqnarray}
which is precisely the equation of continuity for electromagnetic energy density. Now having derived
the BI theory version of the Poynting vector as $\vec{S} = \vec{E} \times \vec{H} = \vec{E} \times (\vec{B}/R)$,
the angular momentum of the electromagnetic field is obtained by integrating the moment of the Poynting
vector over all space which yields
\begin{eqnarray}
\vec{J}_{em} &=& \int d^3x \left[ \vec{x}\times (\vec{E}\times {1\over R}\vec{B})\right] \\
&=& \int d^3x \left[ \vec{x}\times ({1\over R}\vec{E}\times \vec{B})\right] =
\int d^3x \left[ \vec{x}\times (\vec{D}\times \vec{B})\right] \nonumber \\
&=& - \int d^3x (\nabla \cdot \vec{D})\left[ {g\over 4\pi}\hat{x}\right] = -{eg\over 4\pi}\hat{r}
\nonumber
\end{eqnarray}
where we used $\vec{B} = (g/4\pi r^2)\hat{x}$ and $\nabla \cdot \vec{D} = e \delta^3(\vec{x}-\vec{r})$
representing the configuration in which the static source magnetic monopole is sitting at the
origin while the test electric charge, at some point of time, is at $\vec{r}$.  Thus this completes the
demonstration. 
\\
{\bf IV. Concluding remarks}
\\
To summarise, it is interesting to note that in the context of BI electrodynamics, despite
the inherent lack of dual symmetry in BI equations, when we assume the existence of magnetic
monopoles, interesting new physics emerge such as the exotic behavior of static electric monopole
field and the existence of point-like dyons while the cherished principles such as the Dirac
quantisation condition still hold true without experiencing any modification. \\
Concerning the nature of the present work, a word of caution may be helpful to answer to possible
criticism. That is, one might wonder what exactly distinguishes the present work from the
pile of works on similar subjects in the recent literature [3-9]. As we mentioned at the beginning
of the introduction, the revival of interest in the BI gauge theory was triggered by the
recently fashionable D-brane physics [2]. Indeed in the recent literature, one finds a number
of works discussing dyons in abelian BI gauge theory [5-9]. Some of them use the terminology,
``BIons'' for soliton solutions possessing the properties of these dyons. Although these
dyon solutions are also static solutions in abelian BI gauge theory, they all arise in
theories resulted from the dimensional reduction of some higher-dimensional (10-dimensional,
to be more specific) supersymmetric pure abelian BI theory. Being theories which emerge as a
result of dimensional reduction, they inevitably involve one or more scalar fields degrees of
freedom representing the compactified extra dimensions in addition to the 4-dimensional
abelian BI gauge field. And it is precisely these additional scalar fields which play the role
of Higgs-type field in the familiar Yang-Mills-Higgs theory [10] and thus lead to Bogomol'nyi-type
first-order equations [11] of which the solitonic solutions are generally dyon solutions. Thus
the dyon solutions in these brane-inspired theories are really Julia-Zee-type dyon solutions [12]
in nature and the abelian BI gauge field involved behaves as the abelian projection of the
non-abelian Yang-Mills field after the spontaneous symmetry breaking. And the dyon solutions
there rely, for their existence, entirely on the non-vanishing scalar fields having some
particular solution behavior. In addition, the electric charge and the magnetic charge there
are not two independent parameters. Instead, they are generated from a single parameter of the
theory. Besides, it seems worth mentioning that even some early works [3,4] (but in modern perspective)
on 4-dimensional nonlinear electrodynamics, such as that of BI, never considered the physics in the 
presence of magnetic monopoles and discussed only the exotic behavior of static electric monopole field. 
In contrast, our philosophy in the present work was the parallel comparison of monopole physics 
between in standard Maxwell electrodynamics and in the original version of abelian BI electrodynamics 
having no connection whatsoever to the brane physics. Thus the relevant degree of freedom of the theory
is just the abelian gauge field alone and then we discovered static solutions possessing all the evidences 
in favor of the dyon interpretation.
Moreover, this point-like dyon solution carries electric and magnetic charges which are independent of
each other up to Dirac quantisation condition. To conclude, therefore, point-like dyon solution in
the original version of the 4-dimensional BI electrodynamics found in the present work should be
distinguished from the ones with the same name appeared in the recent literature. And as we stressed in the
text, this occurrence of point-like dyon solution in BI electrodynamics can be attributed to the
highly nonlinear nature of the theory, or more precisely, to the unique and direct coupling between
electric and magnetic fields appearing particularly in 4-dimensions even in the static case.
Lastly, in the present work we witnessed that even the simple study of monopole physics exposed some 
of the unique and exciting hidden features of the abelian BI gauge theory and this seems to suggest that
the BI gauge theories, abelian or non-abelian, really deserve serious and full exploration in modern
field theory perspective.

\vspace{1cm}

{\bf \large Acknowledgements}
\\
The author would like to thank Prof. Bum-Hoon Lee for valuable comments. This work was supported
in part by grant of Post-doc. program at Kyungpook National University (1999).

\vspace{1cm}

{\bf \large References}

\begin{description}

\item {[1]} M. Born, Proc. R. Soc. London, {\bf A143}, 410 (1934) ;
            M. Born and M. Infeld,  Proc. R. Soc. London, {\bf A144}, 425 (1934) ;
            P. A. M. Dirac,  Proc. R. Soc. London, {\bf A268}, 57 (1962). 
\item {[2]} J. Polchinski, {\it TASI Lectures on D-branes}, hep-th/9611050 ;
            R. Argurio, {\it Brane physics in M-theory}, hep-th/9807171 ;
            K. G. Savvidy, {\it Born-Infeld action in string theory}, hep-th/9906075,
            and references therein.
\item {[3]} G. W. Gibbons, Nucl. Phys. {\bf B514}, 603 (1998).
\item {[4]} G. W. Gibbons and D. A. Rasheed,  Nucl. Phys. {\bf B454}, 185 (1995).
\item {[5]} C. G. Callan and J. M. Maldacena, Nucl. Phys. {\bf B513}, 198 (1998).
\item {[6]} J. P. Gauntlett, C. Koehl, D. Mateos, P. K. Townsend and M. Zamaklar,
            hep-th/9903156 ; R. de Mello Koch, A. Paulin-Campbell and 
            J. P. Rodrigues, hep-th/9903207.
\item {[7]} D. Bak, J. Lee and H. Min, Phys. Rev. {\bf D59}, 045011 (1999).
\item {[8]} D. Youm, hep-th/9905155.
\item {[9]} D. Brecher, hep-th/9804180.
\item {[10]} G. 'tHooft, Nucl. Phys. {\bf B79}, 276 (1974) ;
             A. M. Polyakov, JETP Lett. {\bf 20}, 194 (1974).
\item {[11]} M. K. Prasad and C. M. Sommerfield, Phys. Rev. Lett. {\bf 35}, 760 (1975) ;
             E. B. Bogomol'nyi, Sov. J. Nucl. Phys. {\bf 24}, 449 (1976).
\item {[12]} B. Julia and A. Zee, Phys. Rev. {\bf D11}, 2227 (1975).

\end{description}

\end{document}